# Astro2020 Project White Paper

# ATLAS Probe: Breakthrough Science of Galaxy Evolution, Cosmology, Milky Way, and the Solar System

**Lead Author:**

Name: Yun Wang
Institution: California Institute of technology
Email: wang@ipac.caltech.edu
Phone: (626) 395-1415

**Co-authors:**

Mark Dickinson (NOAO), Lynne Hillenbrand (Caltech), Massimo Robberto (STScI & JHU), Lee Armus (Caltech/IPAC), Mario Ballardini (Western Cape, South Africa), Robert Barkhouser (Johns Hopkins Univ.), James Bartlett (JPL), Peter Behroozi (Univ. of Arizona), Robert A. Benjamin (Univ. of Wisconsin Whitewater), Jarle Brinchmann (Porto, Portugal; Leiden, Netherlands), Ranga-Ram Chary (Caltech/IPAC), Chia-Hsun Chuang (Stanford), Andrea Cimatti (Univ. of Bologna, Italy), Charlie Conroy (Harvard), Robert Content (Australia Astronomical Obs.), Emanuele Daddi (CEA, France), Megan Donahue (Michigan State Univ.), Olivier Dore (JPL), Peter Eisenhardt (JPL), Henry C. Ferguson (STScI), Andreas Faisst (Caltech/IPAC), Wesley C. Fraser (Queen's Univ. Belfast, U.K.), Karl Glazebrook (Swinburne Univ. of Technology, Australia), Varoujan Gorjian (JPL), George Helou (Caltech/IPAC), Christopher M. Hirata (Ohio State Univ.) Michael Hudson (Univ. of Waterloo, Canada), J. Davy Kirkpatrick (Caltech/IPAC), Sangeeta Malhotra (GSFC), Simona Mei (Paris Obs., Univ. of Paris D. Diderot, France), Lauro Moscardini (Univ. of Bologna, Italy), Jeffrey A. Newman (Univ. of Pittsburgh), Zoran Ninkov (Rochester Insti. of Tech.), Alvaro Orsi (CEFCA, Spain), Michael Ressler (JPL), James Rhoads (GSFC), Jason Rhodes (JPL), Russell Ryan (STScI), Lado Samushia (Kansas State Univ.), Claudia Scarlata (Univ. of Minnesota), Daniel Scolnic (Duke Univ.), Michael Seiffert (JPL), Alice Shapley (UCLA), Stephen Smee (Johns Hopkins Univ.), Francesco Valentino (Univ. of Copenhagen, Denmark), Dmitry Vorobiev (Univ. of Colorado), Risa H. Wechsler (Stanford)

**Type of Activity:** Space-based Project

# I. *ATLAS* Probe: Key Science Goals and Objectives

The observational data from recent years have greatly improved our understanding of the Universe. However, we are far from understanding how galaxies form and develop in the context of an evolving "cosmic web" of dark matter, gas and stars, and the nature of dark energy remains a profound mystery 20 years after the discovery of cosmic acceleration. Understanding galaxy evolution in the context of large-scale structure is of critical importance in our quest to discover how the Universe works. This requires very large spectroscopic surveys at high redshifts: very large numbers of galaxies over large co-moving volumes for robust statistics in small redshift bins ranging over most of cosmic history. In particular, we need to map the cosmic web of dark matter using galaxies through most of cosmic history. This requires a redshift precision of ~0.0001 (i.e., R=1000 slit spectroscopy), and continuous IR wavelength coverage only possible from space. These observational requirements also enable definitive measurements on dark energy with minimal observational systematics by design. A very high number density wide area galaxy redshift survey (GRS) spanning the redshift range of 0.5<$z$<4 using the same tracer, carried out using massively parallel wide field multi-object slit spectroscopy from space, will provide definitive measurements that can illuminate the nature of dark energy, and lead to revolutionary advances in particle physics and cosmology. The currently planned projects do not meet these science goals. *JWST* has slit spectroscopic capability, but a relatively small Field of View (FoV), thus unsuitable for carrying out surveys large enough to probe the relation between galaxy evolution and environment in a statistically robust manner. Both *Euclid* and *WFIRST* employ slitless grism spectroscopy, which increases background noise, and only cover wavelengths below 2μm with fairly low spectral resolution, both of which will limit their capability to probe galaxy evolution science. While *Euclid* and *WFIRST*, and the ground-based projects (e.g., *DESI, PFS*, and *LSST)*, will significantly advance our understanding of the nature of dark energy, they will not provide definitive measurements for its resolution, due to limits inherent to each (see e.g., Wang et al. 2019b). **The lack of slit spectroscopy from space over a wide FoV in the Near and Mid IR is the obvious gap in current and planned future space missions. *ATLAS* fills this gap in order to address the fundamental questions on galaxy evolution and the dark Universe.**

*ATLAS* (Astrophysics Telescope for Large Area Spectroscopy) is a concept for a NASA probe-class space mission that will achieve groundbreaking science in all areas of astrophysics. It is the spectroscopic follow-up space mission to *WFIRST*, boosting its scientific return by obtaining deep 1-4μm slit spectroscopy in three tiered galaxy redshift surveys (wide: 2000 deg$^2$; medium: 100 deg$^2$; deep: 1 deg$^2$) for most of the galaxies imaged by the ~2000 deg$^2$ *WFIRST* High Latitude Survey (HLS) at $z$>0.5. *ATLAS* spectroscopy will measure accurate and precise redshifts for ~200M galaxies out to $z$=7 and beyond, and deliver spectra that enable a wide range of diagnostic studies of the physical properties of galaxies over most of cosmic history. *ATLAS* and *WFIRST* together will produce a definitive 3D map of the Universe over 2000 deg$^2$ (Fig.1).

***ATLAS* Probe Science Goals** are: (1) Discover how galaxies have evolved in the cosmic web of dark matter from cosmic dawn through the peak era of galaxy assembly. (2) Discover the nature of cosmic acceleration. (3) Probe the Milky Way's dust-enshrouded regions, reaching the far side of our Galaxy. (4)

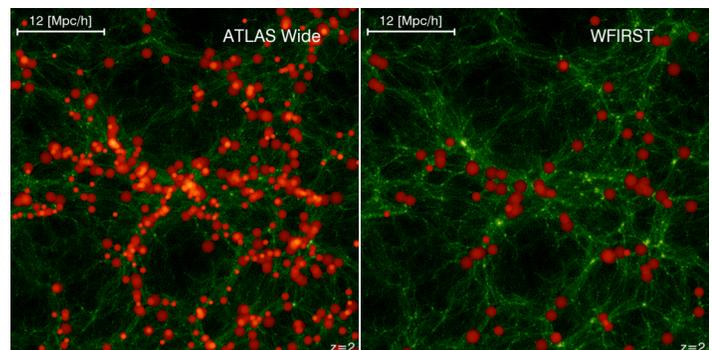

Fig. 1: Cosmic web of dark matter (green) at z=2 traced by galaxies (red filled circles) from the *ATLAS* Wide survey (left), which obtains spectra for 70% of galaxies in the *WFIRST* weak lensing sample, compared to *WFIRST* GRS (right). The larger circles represent brighter galaxies. (Wang et al. 2019a)

Discover the bulk compositional building blocks of planetesimals formed in the outer Solar System. These flow down to the **ATLAS Probe Scientific Objectives**: (1A) Trace the relation between galaxies and dark matter with less than 10% shot noise on relevant scales at $1<z<7$. (1B) Probe the physics of galaxy evolution at $1<z<7$. (2) Obtain definitive measurements of dark energy and tests of General Relativity. (3) Measure the 3D structure and stellar content of the inner Milky Way to a distance of 25 kpc. (4) Detect and quantify the composition of 3,000 planetesimals in the outer Solar System.

*ATLAS* is a 1.5m telescope with a FoV of 0.4 deg$^2$, and uses Digital Micro-mirror Devices (DMDs) as slit selectors. It has a spectroscopic resolution of R = 1000, and a wavelength range of 1-4 $\mu$m. *ATLAS* has an unprecedented spectroscopic capability based on DMDs, with a spectroscopic multiplex factor ~6,000. *ATLAS* is designed to fit within the NASA probe-class space mission cost envelope; it has a single instrument, a telescope aperture that allows for a lighter launch vehicle, and mature technology (DMDs can reach TRL 6 within two years). The pathfinder for *ATLAS*, *ISCEA* (Infrared SmallSat for Cluster Evolution Astrophysics), has been selected by NASA for a mission concept study. We anticipate *ATLAS* to be launch ready by 2030. *ATLAS* will lead to transformative science over the entire range of astrophysics. We will briefly summarize *ATLAS* science below. **Wang et al. (2019a) presents ATLAS Probe in detail. Astro2020 science white papers by Behroozi et al. (2019), Dickinson et al. (2019), Pisani et al. (2019), and Wang et al. (2019b) address ATLAS science in galaxy evolution and cosmology.**

**(i) Decoding Galaxy Evolution Physics Using *ATLAS* Probe**

In today's era of precision cosmology, we believe that we understand the growth of structure in a universe of cold dark matter and dark energy, and we can map this over cosmic time with sophisticated numerical simulations. However, the galaxies that we see are not simply dark matter halos. Baryonic physics makes them far more complex, and we are still far from understanding how galaxies form and develop in the evolving context of large scale structure. *ATLAS* Probe will provide wide field, highly multiplexed, densely sampled spectroscopy at high redshifts, producing a "time-lapse *SDSS*" spanning most of cosmic history. *ATLAS* will carry out three nested surveys (Wide/Medium/Deep): 2000 deg$^2$ to the line flux limit of $5\times10^{-18}$ erg/s/cm$^2$; 100 deg$^2$ to AB~25; 1 deg$^2$ to AB~26. These will sample the galaxy luminosity function over a wide range of redshifts. Spectroscopic detection of H$\alpha$ (among other lines) out to $z=5$, and [OIII]+H$\beta$ in the late reionization era, to $z=7$, will relate each galaxy to its place in the cosmic web with precision that cannot be achieved with photometric redshifts or slitless spectroscopy (Fig.2). Detection of multiple emission lines will provide diagnostics of ISM excitation, metal abundance and dust reddening. The Medium and Deep surveys will also measure absorption line redshifts for quiescent galaxies.

Galaxy properties correlate with those of the underlying dark matter halos: their masses, spins, positions, and environments. *ATLAS* surveys will allow statistical derivation of dark matter halo masses from clustering of galaxies binned by other observable/inferable parameters such as stellar mass, star formation rate or morphology. This information will be used to derive the *stellar mass—halo mass relationship* (SMHMR, Moster et al. 2013, Behroozi et al. 2013), which measures the efficiency with which galaxies turn gas into stars, and is a key probe of the strength of feedback from stars and supermassive black holes. *ATLAS* will measure the SMHMR as a function of galaxy properties and its evolution over cosmic time. Additional constraints will come from spectroscopic *group catalogs* and *local environmental density*. *ATLAS* can also measure average dark matter accretion rates for galaxies via the detection of the splash-back radius (a.k.a., turnaround radius) of their satellites (More et al. 2016) out to $z \sim 5$.

The shape of the SMHMR implies a characteristic mass at which galaxies most effectively convert gas into stars. At lower and higher masses, "feedback" is invoked to prevent gas from cooling onto galaxies, or to expel gas, thus suppressing star formation efficiency. Supermassive black holes power active galactic nuclei (AGN), which may regulate star formation and growth. *ATLAS* spectroscopy will identify vast samples of high-redshift AGN using standard nebular excitation diagnostics (e.g., Baldwin, Phillips, &

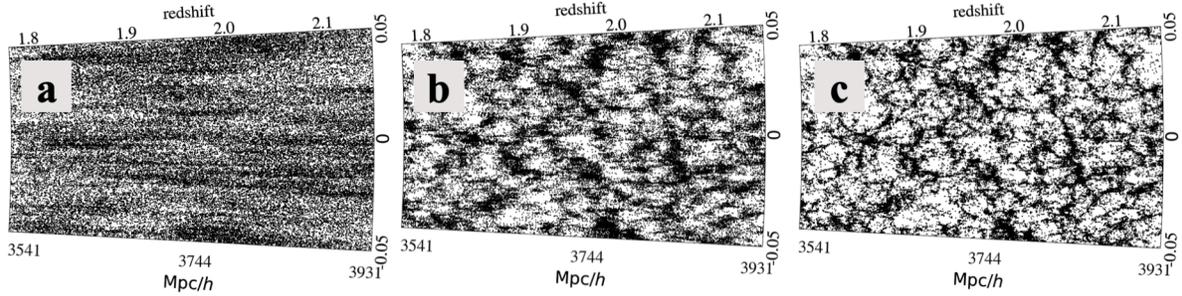

Fig.2: The spatial distribution of Hα-emitting galaxies at z=2 from the semi-analytical galaxy formation model GALFORM. Each panel illustrates a different survey of the same galaxy distribution, with redshift accuracy $\sigma_z/(1+z)$ equal to (a) $10^{-2}$ (most optimistic photo-zs); (b) $10^{-3}$ (slitless spectroscopy); and (c) $10^{-4}$ (*ATLAS* slit spectroscopy).

Terlevich 1981). [OIII] luminosities will provide a measure of the accretion luminosities and black hole growth rates. *ATLAS* will connect AGN activity to local and large-scale environment with exquisite statistical accuracy that is only possible today in the local universe.

Current observations indicate that the intergalactic medium (IGM) completed its transition from neutral to ionized around z ~ 6, but the processes responsible are poorly understood. Reionization may have been highly inhomogeneous, with expanding bubbles driven by strongly clustered young galaxies that are highly biased tracers of dark matter structure. Future radio facilities (LOFAR, HERA, SKA, ASKAP) will map (at least statistically) the distribution of neutral hydrogen in the epoch of reionization. *ATLAS* surveys will provide complementary maps of the spatial distribution of the (potentially) ionizing galaxies themselves over the same volumes, detecting [OIII]+Hβ emission lines at 5<z<7 over very wide sky areas. An accurate measurement of 3D clustering will strongly constrain theoretical models that can then be extrapolated to higher redshifts, earlier in the epoch of reionization. There is already evidence that Lyα may be inhomogeneously suppressed by the neutral IGM at z > 7, (e.g., Tilvi et al. 2014), and that its escape may correlate with galaxy overdensities that can more effectively ionize large volumes (Castellano et al. 2016). *ATLAS* spectroscopy can detect or set limits on Lyα emission from vast numbers of galaxies selected photometrically from deep *WFIRST* surveys, providing additional constraints on the reionization process.

The *ATLAS* surveys will enable a wide range of additional investigations. They will provide spectroscopic identification and confirmation for hundreds of massive galaxy clusters at "cosmic noon" (z ≈ 2), as well as early groups and proto-clusters in the *ATLAS* deep survey out to z ≈ 7. Galaxy emission line widths can be interpreted in concert with *WFIRST* imaging and structural properties to infer galaxy velocity functions. Velocity shifts between ISM absorption lines (e.g., MgII 2800Å, NaI 5890,5896Å) and galaxy systemic redshifts (e.g., from Hα or [OIII] emission) can be used to trace gas flows around star-forming galaxies. Densely-sampled spectroscopy can also be used to study galaxy pairs and the evolution of the merger fraction and merger rate. Overall, *ATLAS* will provide rich and abundant spectroscopy to fully exploit the wealth of information from the *WFIRST*, *Euclid* and *LSST* imaging surveys, firmly connecting hundreds of millions of galaxies to the evolving cosmic web.

### (ii) Definitive Measurements of Dark Energy from *ATLAS* Probe

Given our ignorance of the nature of dark energy, it is critical that we obtain measurements on dark energy that are model-independent (cosmic expansion history $H(z)$ & growth of large scale structure $f_g(z)$ as free functions) and definitive (high precision and accuracy) over the entire redshift range over which dark energy influences the expansion of the Universe (i.e., 0<z<4). ATLAS Wide covers 2000 deg$^2$ at 0.5<z<4, with a galaxy surface number density ~12 times that of the *WFIRST* GRS and ~50 times that of *Euclid*, with spectroscopic redshifts for 183M galaxies (see Fig.1). **Galaxy clustering data from 3D distributions of galaxies is the most robust probe of cosmic acceleration.** The baryon acoustic oscillation (BAO) measurements provide a direct measurement of $H(z)$ and angular diameter distance $D_A(z)$ (Blake & Glazebrook 2003; Seo & Eisenstein 2003), and the redshift-space distortions (RSD) enable measurement of $f_g(z)$ (Guzzo et al. 2008; Wang 2008). *ATLAS* Wide provides multiple galaxy tracers of BAO/RSD (red galaxies, different emission-line selected galaxies, and WL shear selected galaxies) over 0.5<$z$<4, with each at high number densities. These enable robust modeling of BAO/RSD (e.g., the removal of the nonlinear effects via the reconstruction of the linear density field), and significantly tightens constraints on dark energy and modified gravity by evading the cosmic variance when used as multi-tracers (McDonald & Seljak 2009). *ATLAS* Wide enables detailed study of the galaxy-formation systematics (feedback; assembly bias; conformity) that are potential systematics in modeling RSD (Tojeiro et al., 2017). It measures $H(z)$, $D_A(z)$, and $f_g(z)$ over the wide redshift range of 0.5<$z$<4 (see Fig.3), with high precision over 0.5<$z$<3.5. If early dark energy remains viable in the 2020s, it can be measured by enhancing *ATLAS* Wide with a high z survey targeting Hα emission line galaxies at 3<$z$<4 selected from *WFIRST* HLS imaging.

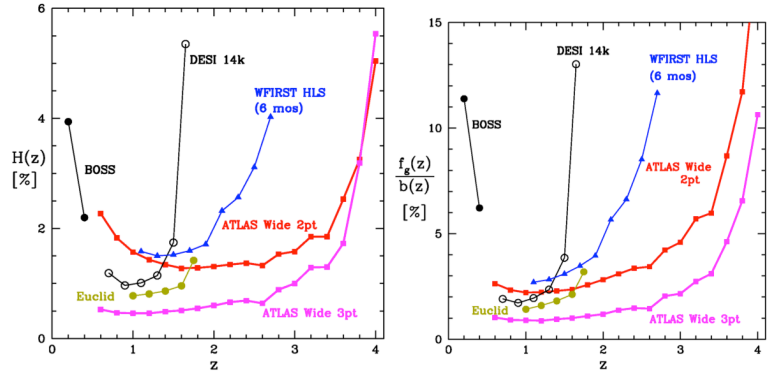

Fig.3. Expected $H(z)$ and $f_g(z)$ from future surveys. "2pt" refers to galaxy power spectrum, "3pt" refers to galaxy bispectrum. Constraints are derived following Wang et al. (2013) & Samushia et al. (2019). The constraints on $D_A(z)$ (not shown to avoid cluttering) provide a cross-check on $H(z)$. The bias between galaxy and matter distributions is $b(z)$. *ATLAS* overlaps ground-based projects 0.5<$z$≲1 for key cross-check and mitigation of systematic effects).

The very high number density galaxy samples from the *ATLAS* Wide survey provide the ideal data set for studying higher-order statistics of galaxy clustering. For a galaxy sample with number density $n$, shot noise scales as 1/$n$ for 2pt, and 1/$n^2$ for 3pt statistics. Fig.2 shows that *ATLAS Wide 3pt statistics gives definitive measurements on dark energy, outperforming all other measurements* (Samushia et al. 2019). Since the 3pt statistics provides information not contained in the 2pt statistics, the combination of these is needed to optimally extract the cosmological information from galaxy clustering data (see, e.g., Gagrani & Samushia 2017), and enables the direct measurement of bias $b(z)$. In addition, the cross-3pt function, galaxy-galaxy-lensing shear, will help break degeneracies between galaxy bias and cosmological parameters. It will be measured with high signal-to-noise for the sample sizes discussed here. While the use of galaxy clustering 2pt statistics is now standard in cosmology, the use of the 3pt statistics is still limited due to a number of technical challenges (see, e.g., Yankelevich & Porciani 2019). *ATLAS Wide in the next decade will take advantage of the anticipated future advances in galaxy 3pt statistics to deliver game-changing science.*

**(iii) Probing the Dust-Obscured Inner Milky Way With *ATLAS* Probe**

The *ATLAS* Wide Survey at high Galactic latitude will also probe the foreground stellar content of the Milky Way to unprecedented spectroscopic depth. In addition, a dedicated *ATLAS* Galactic Plane Survey will unveil and characterize objects at all evolutionary phases, from deeply embedded class-0 protostars to the most elusive dusty Luminous Blue Variables. With SNR>30 spectra for 95M sources having AB<18.2 mag, and SNR>5 to AB=21.5 for hundreds of millions more stars, covering 700 deg$^2$ in 0.4 years of observing time, this survey will advance our understanding of the structure, star-forming history, and stellar content of the Milky Way.

Currently, we know more about the structure and the spatially resolved star formation histories of galaxies in and even beyond the Local Group, than we do about our own Galaxy. *ATLAS* will advance our understanding of the 3D Milky Way beyond the ongoing revolution provided by ESA/Gaia, especially in the inner Galaxy ($|l|<65°$ and $|b|<1°$) where 98.5% of the Galactic plane has G-band extinction >7.5 mag. Going to the infrared drops the high extinction fraction to only 10.4% of the Galactic plane in the K band. *ATLAS*' infrared range thus opens up large regions of the Galaxy for spectral investigation. *ATLAS* will cover both the inner and outer Galaxy, and thus very different regions in terms of galactic structure, star formation, evolved star properties, and interstellar dust.

*ATLAS* spectroscopy will follow on the advances of *2MASS*, *WISE*, and *Spitzer* in identifying candidates of particularly interesting object classes based on colors alone, and the expected return from *WFIRST*. For *WFIRST* photometry in particular, stellar photospheric temperatures and interstellar extinction effects will be almost completely degenerate due to the filter choices. *ATLAS* spectra will also complement the planned *SDSS*-V spectroscopic survey programs which will cover brighter Milky Way objects ($H<11.5$ mag).

For **Galactic Structure**, *ATLAS* spectroscopy can produce unique information on the bar(s) of the Galaxy, the nuclear region, the stellar disk, and spiral arms, relative to existing and planned photometric surveys that use star counts techniques and statistical analysis of color-color and color magnitude diagrams. Not only will *ATLAS* be able to provide spectral information on all of the (nearby) stars observed in the optical by *Gaia*, it will be able to see much of the substantial stellar population that is not detectable in the optical. Reconstructing the 3D structure of the Galaxy will allow progress on a number of fundamental questions regarding e.g. the scale-length of the Galactic disk, whether the stellar warp and the gas warp coincide, and the existence of stellar streams across the Galactic plane.

For **Star Formation,** the *Spitzer*/GLIMPSE surveys produced an unprecedented picture of star formation in our Galaxy by unveiling hundreds of new star forming regions. *ATLAS* will quantify the Star Formation Rate (SFR) of the Galaxy, its variation with Galactocentric radius, and its association with various dynamical features in the Galaxy, testing theories of star formation both on a global scale and at the molecular cloud level. *ATLAS* will characterize Young Stellar Objects (YSO) and their surrounding environments, analyzing the energy budget and mapping the spatially resolved star formation history of the Galaxy over the past <50-100 Myr.

For **Interstellar Extinction**, by measuring the extinction law over tens of millions of lines of sight, *ATLAS* will provide a unique dataset to study in detail the variation of the extinction curve out to the edge of the Galactic disk. Variations with Galactic longitude have been identified and attributed to small variations in ISM density, mean grain size, or disk metallicity gradient. Variations of chemical composition of dust grains reflect the abundance/depletion of metals in the ISM, and hence the cooling mechanisms that control the efficiency of star formation.

For **Substellar Objects**, *ATLAS* will study brown dwarfs as they cool through the L-T-Y spectral sequence with age. These objects provide important information on the shape and low-mass cutoff of the field mass function. They also stand as proxies for exoplanets, given their atmospheres of similar effective temperatures. At the $3\sigma$ limit of $J\approx23.5$ mag AB, *ATLAS* will detect 5-Gyr-old field brown dwarfs as low in mass as 10 MJup (Teff~300K, an early-Y dwarf) at 10 pc and 35 MJup (Teff~700K, late-T) at 100 pc.

For **Evolved Stars,** *ATLAS* will detect Red Giant Clump star standard candles with $L$mag=$-1.75$ (0.71 in AB mag) at 10kpc for $Av = 30$ mag, reaching the heavily obscured regions of the Galactic Center or the outer edges of the Milky Way over at least the two outer quadrants. Red Supergiants can be studied across the entire Galactic disk. AGB stars (initial mass 4-8$M_\oplus$) are sufficiently short lived that they can be used to trace the spiral arms of the Galaxy. Their number and distribution provides a fossil record of the recent history of star formation in the Milky Way.

### (iv) Exploring the Outer Solar System With *ATLAS* Probe

Despite more than 2 decades of spectroscopic observations of Kuiper Belt Objects (KBOs), very little is known about their surface compositions; beyond water-ice and methanol, no materials have been confidently identified in the spectra of small (D≲500km) KBOs. This is largely the result of the lack of identifying absorption features in the λ≲2.5μm region, beyond which current facilities are insufficiently sensitive to gather observations of these bodies. This is unfortunate, as many anticipated materials exhibit strong absorption features at these longer IR wavelengths (Parker et al., 2016).

Notably, no silicate materials, commonly identified by Fe/Mg absorptions at ~1μm, and deep hydroxl feature at 3.0μm, have ever been detected in the spectra of KBOs. This represents one of the big outstanding gaps in our compositional knowledge of these bodies. From our best proxy of the spectra of KBOs - Phoebe - it is clear that *ATLAS* Probe has the potential to detect such features, as it will provide the requisite SNR throughout the important 1-4μm wavelength range.

Spectra of KBOs will come from two separate *ATLAS* surveys. The *ATLAS* Wide Survey will gather useful IR reflectance spectra of roughly 300 KBOs. These spectra will provide NIR spectral slopes, as well as a measure of water-ice absorption. For the brightest targets, silicate detection is possible. The *ATLAS* Solar System Survey, with pointed observations of known bodies, would gather more than 3000 spectra, down to a practical brightness limit of *r*~23.2, for on-target integrations of 2500s. All such bodies have, or will be detected and tracked by the *LSST*, or the *Pan-STARRS* surveys, and the resultant spectra will have higher SNR than typical from the spectra gathered during the ATLAS Wide Survey.

## II. *ATLAS* Probe: Technical Overview

To meet its science objectives, *ATLAS* requires a ~1.5m space telescope in an L2 orbit (Fig.4), with a multi-object spectrograph with R~1000 over a FoV ~0.4 deg$^2$, with spectroscopic multiplex of ~6000, and the wavelength coverage of 1-4μm.

### (i) *ATLAS* Probe Instrument

**Optical Design:** *ATLAS* has only one instrument consisting of 4 identical modules, compact and modular; it fits below the primary mirror structure into a cylindrical envelope only slightly larger than 1.5m in diameter (the size of the primary) and ~65cm in height (see Fig.5). The instrument size can be reduced in a future design phase. The camera optics image a square 0.75″ field onto a little less than 2 x 2 pixels on the detector, delivering a scale of about 0.385″ pixel. The footprint of the DMD on each NIR detector is about 4,000 x 2,100 pixels. The scale can easily be changed in the spectral direction to at least 2.1 pixel/micro-mirror because of the unused space on the 4k×4k detector. The instrument is maintained at temperatures

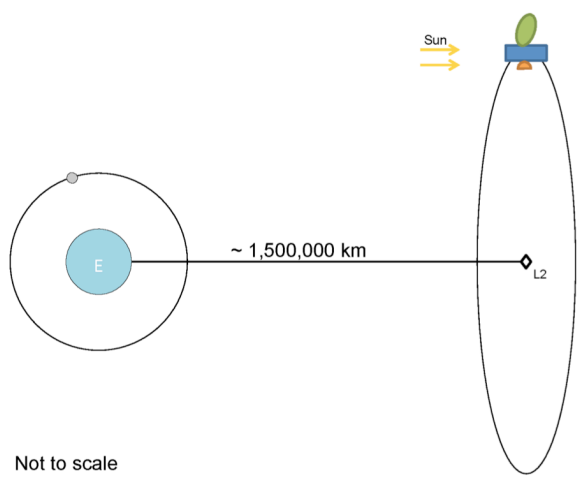

Fig.4: *ATLAS* Probe orbit.

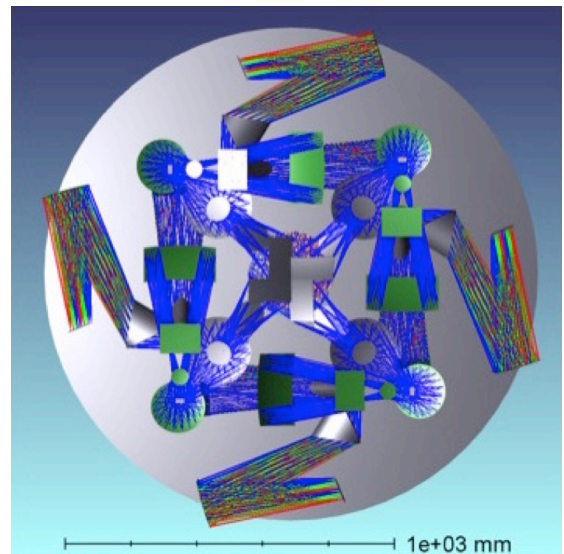

Fig.5: A full view of the preliminary optical design for the *ATLAS* Probe instrument. The large gray circle is the back of the primary.

~50K to keep thermal noise below zodiacal light level. Our preliminary optical design already has excellent image quality in the spectrograph. The Gaussian Equivalent Full Width at Half Maximum (GEFWHM) is about 1/2 of a DMD micro-mirror image. The image quality of the fore-optics on the DMD is 1/2 of a micro-mirror, satisfactory at this point of the preliminary design. Table 1 lists the main parameters of our system. For a detailed discussion of the *ATLAS* instrument, see Wang et al. (2019a).

**Target Selection Mechanism:** The *ATLAS* instrument requirement of a spectroscopic multiplex factor of ~6000 drives the adoption of DMDs as the target selection mechanism. DMDs have been invented for digital display/projection applications by Texas Instruments. *ATLAS* baselines the 2k CINEMA model with 2048×1080 micro-mirrors, 13.7μm on a side. Each micro-mirror of the DMD's can tilt ±12° to separate the reflected "ON" vs. "OFF" beam. Therefore a DMD, perpendicularly illuminated, must receive a beam slower than f/2.4 to prevent overlap between the input and output beams, setting an upper limit to the scale per micro-mirror, and therefore to the total FoV of the spectrograph.

| TELESCOPE | |
|---|---|
| Type | modified Ritchey-Chrétien |
| Primary: diameter & focal ratio | 150cm; f/1.6 |
| Primary: central obscuration | 19% diameter (3.7% area) |
| Secondary: diameter | 29 cm |
| Telescope focal ratio | f/11.2 |
| PYRAMID MIRROR | 4 rectangular faces |
| Size | 4 ×13.6 cm×7.4 cm |
| Field of view | 4 ×25.60′×13.50′ |
| FORE OPTICS | |
| f/# (off axis) | f/2.3×f/2.5 |
| Scale on DMD (slit size) | 0.75″×0.75″/micro-mirror |
| COLLIMATOR | |
| Elements | 4 mirrors (+ 1 dichroic) |
| DISPERSING ELEMENTS | |
| Wavelength ranges | 1-2.1μm (NIR); 2.1-4μm (MIR) |
| Resolving power | R ~ 1000 |
| Type | prism |
| CAMERAS | |
| Elements | 1 mirror (each camera) |
| Sampling on detector | 0.38″×0.39″ (pixel scale) |

Table 1: *ATLAS* main optical parameters.

**Detectors and ASICs:** Our baseline detector is the Teledyne H4RG-10, the same type currently under development for *WFIRST*. The long-wavelength cutoffs of our spectroscopic channels (2.1μm and 4μm) are compatable with the standard ~2.35μm and ~5.37μm cutoff of *WFIRST* and *JWST* devices respectively. Fine-tuning the Hg vs. Cd stoichiometric ratio can further reduce the long wavelength cutoff and therefore the dark current, allowing warmer operating temperatures. H4RG arrays tested by *WFIRST* have typical QE>90% in the 0.8-2.35μm range, readout noise ~15e in Double Correlated Mode and mean dark current <0.01e/s/pixel. *JWST* devices have similar performance. The architecture of the multiplexer allows multiple non-destructive reads of each pixel during a single exposure ("sampling up the ramp"); this enables mitigation against cosmic rays and reduction in read noise: typical readout noise with 16 non-destructive samples drops to about 5e. Teledyne has developed SIDECAR ASICs (Application Specific Integrated Circuit) to manage all aspects of FPA operation and output digitization in cold environment. By keeping analog signal paths as short as possible to reduce noise and output capacitance loading, ASICS improve power consumption, speed, weight and performance. A Teledyne ASIC device is currently driving *HST*/ACS and several Teledyne ASICs will soon fly on 3 out of 4 *JWST* instruments; we will adopt such devices for *ATLAS*.

**(ii) *ATLAS* Probe Mission Architecture:**

**Mission Implementation:** We assume a 5 year *ATLAS* mission in a halo L2 orbit similar to that of *JWST* (Fig.4), enabling long observations in a very stable thermal and radiation environment; the narrow Sun-Spacecraft-Earth angle facilitates passive cooling to the ~50K operating temperature needed for the long wavelength channel detectors. Absolute pointing requirements are relaxed due to the versatility of the DMDs; this results in a pointing stability requirement of 0.1″ over the 1000s exposure time which is adequate for the 0.39″ pixels. The temperature of the instrument is low enough to make it immune to changes related to the spacecraft attitude. Note that for a survey telescope the attitude can be more easily maintained within optimal range than for a general observatory (like e.g. *HST* or *JWST*) where the pointing and orientation are driven by the science programs, impacting optimal scheduling. Table 2 summarizes *ATLAS* mission implementation requirements. We have assumed that solar panels can be integrated with bottom sun shield / bottom deck of spacecraft. Thermal solutions adopted for similar class missions are adequate for *ATLAS*'s 4 spectrometers.

**Mass Estimates**: *ATLAS* instrument mass were derived by scaling results from studies of similar class missions. A single spectrometer mass and costs is scaled by the number of detectors, for a total of 4 identical spectrometers using 2 H4RG detectors each. The focal plane mass/costs are scaled from a similar study with H2RG detectors. Rules of thumb were applied to calculate spectrometer mass and costs from focal plane mass/cost. The telescope mass was calculated from telescope diameter, using a linear fit derived from similar study results to estimate mass as a function of diameter. The error in the fit was used to calculate max and min probable mass. The payload mass is given as the sum of 4 spectrometers plus the telescope, with estimated minimum, mode, and maximum payload mass, see Table 3.

*ATLAS* spacecraft mass estimate is based on scaling of similar missions using the total payload mass (Telescope + 4 Spectrometers), see Table 4. Estimated minimum, mode, and maximum mass are given, as well as 70th percentile confidence spacecraft mass. Payload mass estimates are used to generate spacecraft bus estimates using historical average Mass/Payload ratios. Bus cost estimates are generated using another relationship between astrophysics spacecraft mass to cost ratios. The bus costs have been increased by $10M to accommodate out-of-family pointing requirements.

| Mission | • Astrophysics IR all sky observer (galaxy redshift survey)<br>• L2 Orbit<br>• Class B Mission<br>• Dual string spacecraft bus |
|---|---|
| Constraints | • Tight pointing stability<br>    – Driven by slit size of 0.75"<br>    – Requires +/- 0.375" 3-sigma<br>• ~40-60 K detector temperature (in family with other passively cooled similar missions)<br>• Sunlight cannot contact telescope<br>• Long exposures (up to days) |
| Measurement | • 4 identical spectrometers covering 1-4 $\mu$m |
| Data Volumes | • ~600Mb every 500 seconds, for 170 samples per day<br>• ~186 Tb over 5 years |
| Commanding | Weekly commanding cycle once on orbit. |

Table 2:. *ATLAS* Probe mission implementation requirements.

| Detectors (#) | Focal Plane mass (kg) | Spectrometer min mass (kg) | Spectrometer mode mass (kg) | Spectrometer max mass (kg) |
|---|---|---|---|---|
| 2 | 2.4 | 26.6 | 39.9 | 79.7 |
| **Telescope** | Diameter (m) | Telescope min mass (kg) | Telescope mode mass (kg) | Telescope max mass (kg) |
| | 1.5 | 214.8 | 350.1 | 485.5 |
| **Payload** | | Payload min mass (kg) | Payload mode mass (kg) | Payload max mass (kg) |
| | | 321.1 | 509.6 | 804.4 |

Table 3: Estimated instrument mass for *ATLAS* Probe.

| | Payload Mass Estimate | Bus Mass Estimate | Bus Cost Estimate |
|---|---|---|---|
| Minimum Payload Mass | 321 kg | 618 kg | $140.05 |
| Mean Payload Mass | 510 kg | 981 kg | $216.41 |
| Maximum Payload Mass | 804 kg | 1548 kg | $335.82 |
| 70% Values | 598 kg | 1150 kg | $252.08 |

| Maximum mass estimate used for margin calculations | EELV Capability (kg) | 3400 | LV margin uses an average L/V adapter mass of 30 kg and propellant mass of 161.7 kg |
|---|---|---|---|
| | Total ATLAS MEV Mass (kg) | 2352 | |
| | NASA Margin (%) | 48% | |

Table 4: Estimated spacecraft mass and cost for *ATLAS* Probe.

**Ground System Design:** The *ATLAS* Ground system is based on a mission specific implementation of the standard JPL mission operations and ground data systems. During Phase E, *ATLAS* spacecraft will cruise to Sun-Earth L2 and enter HALO orbit, and perform surveys per plan to meet science objectives. *ATLAS* will observe most of each day, permitting ~25 minutes/day for miscellaneous spacecraft (S/C) activities that may impact observation (S/C trim maneuvers, wheel desats, antenna pointing…). Science downlink is planned for once a week for 2 hours which can be concurrent with science observations. The ground network utilizes the DSN 34m BWG subnet, the Near Earth Ka-band for High Rate science downlink, 150 Mb/s data rate. Excluding protocol overhead, this becomes effectively 130 Mb/s science downlink. The uplink and engineering data can be X-/S-band via S/C LGA. During cruise, 2-3 passes will be made per week. During science phase, 1-2 passes will be made per week as needed; nominally 1 pass will be adequate. Table 5 shows the Delta V budget for the *ATLAS* Mission Design & Navigation.

| TCM 1 Clean up | 25m/s |
|---|---|
| Halo orbit insertion + Clean up | 25m/s |
| Station Keeping /year | 4m/s |
| **Total over 5 years** | **70m/s** |

Table 5: Delta V budget for *ATLAS* Probe.

## III. *ATLAS* Probe: Technology Drivers

**DMDs and Controller:** The DMD is the core of our system. We base our design on the 2k CINEMA model of 2048×1080 micro-mirrors, 13.7μm on a side. DMDs have been successfully used on ground-based spectrographs like RITMOS (Meyer et al. 2004) and IRMOS (MacKenty et al. 2006). A new DMD-based spectrograph, SAMOS, is under construction for the SOAR telescope in Chile (Robberto et al. 2016). NASA has funded a Strategic Astrophysics Technology (SAT) program (PI Ninkov, *ATLAS* team member) to raise the TRL level of DMDs to TRL5-6 before the 2020 Decadal Survey. DMDs have successfully passed proton and heavy-ion irradiation testing (Fourspring et al. 2013, Travinsky et al. 2016). Following NASA General Environmental Verification Standard (GEVS), the team performed random vibration, sine burst, and mechanical shock testing of manufacturer-sealed DMDs (Vorobiev et al. 2016) and of devices re-windowed for better UV and IR capabilities (Quijada et al. 2016). These GEVS tests suggest that DMDs are robust and insensitive to the potential vibroacoustic environments experienced during launch. Low-temperature testing of DMDs were also performed, the main concern being micro-mirror stiction. Tests at RIT have shown that temperatures as low as 130 K do not affect the performance of DMDs. More recent data obtained at JHU at ~80K confirm these results. A general overview of the results obtained by these test campaigns has been recently presented (Travinsky et al. 2017).

These findings place DMDs between TRL levels 5 and 6 (note that level 6 requires testing against the specific environment of a mission, thus is only achievable after a mission has been identified, e.g., *ATLAS* at L2). DMDs are normally controlled by commercially available boards based on the DLP Discovery 4100 chipset produced by Texas Instruments. The chipset is not designed to operate in a cryogenic environment or in space. *ATLAS* team members at JHU have successfully developed custom electronics to operate Cinema DMDs at cryogenic temperatures. For *ATLAS*, we intend to produce a version of this system based on rad-hard components, suitable for operations at L2.

## IV. *ATLAS* Probe: Organization, Partnerships, and Current Status

The *ATLAS* Probe Collaboration currently has ~50 members, including scientists from the U.S., Italy, France, Spain, Portugal, Canada, Australia, and other countries. The PI is Yun Wang (Caltech). Massimo Robberto (STScI & JHU) is the Instrument Lead. Mark Dickinson (NOAO) and Lynne Hillenbrand (Caltech) are Science Leads. We expect that JPL will be the primary partner for *ATLAS* Probe. JPL funded a high-level TeamX cost study of *ATLAS* probe in June 2019. The pathfinder for *ATLAS*, *ISCEA* (Infrared SmallSat for Cluster Evolution Astrophysics), has been selected by NASA for a mission concept study.

## V. ATLAS Probe: Schedule

Table 6 shows the schedule for *ATLAS* Probe estimated from JPL's Team X schedule reference model, using the closest analog missions. It includes one month schedule reserves for each year in development with 2 months held in ATLO which are fully funded reserves and included in the cost estimate.

## VI. ATLAS Probe: Cost Estimate

The ROM cost estimate for *ATLAS* in Table 7 is based on this methodology: (1) The telescope OTA is from the Stahl Model. (2) The spectrometer(s) cost distribution is scaled from similar class missions and JPL rules of thumb. (3) The spacecraft is based on triangular distribution of Probe class spacecraft busses. (4) Mission Operations and Ground Data System are costed by JPL subsystem engineers. (5) WBS elements 1-4 and 10 are based on their historical relationship to WBS 5 & 6. (6) Standard schedule used for all Probe class missions is applied.

$26.64M (FY19$) total telescope cost was calculated using the 'Phillip Stahl Model'. The focal plane costs are scaled from the number of detectors compared to similar studies. Rules of thumb have been applied to go from focal plane costs to first unit spectrometer cost. The other 3 spectrometers are assumed to be 40% initial unit costs. The focal plane is estimated to cost ~$5.87M, with the 4 spectrometers costing $75.49M, $116.67M, and $213.89M (min, mode, max). The ground system cost is based on Quick Model run based on characteristics of the mission, with development $31M FY19, operations $21M FY19 W/O MD/Nav support. The assumed MD/Nav is ~$7M for development, ~$2.5M for operations.

*The cost information contained in this document is of a budgetary and planning nature and is intended for informational purposes only. It does not constitute a commitment on the part of JPL and/or Caltech.*

| Schedule (months) | |
|---|---|
| Phase A | 12 |
| Phase B | 12 |
| Phase C | 22 |
| Design | 10 |
| Fabrication | 6 |
| Subsystem I&T | 6 |
| Phase D | 18 |
| System I&T | 14 |
| Launch Operations | 4 |
| Phase E | 60 |
| Phase F | 4 |

Table 6: Estimated *ATLAS* Probe schedule.

| WBS No. | WBS Title | Cos Estimate Method | Min | A-D Mode | Max | E-F |
|---|---|---|---|---|---|---|
| 01 | Project Mgmt. | % Wrap from similar studies | $9.5 | $14.0 | $22.5 | $4.3 |
| 02 | Project Sys. Eng. | % Wrap from similar studies | $12.5 | $18.5 | $29.7 | |
| 03 | S&MA | % Wrap from similar studies | $12.9 | $19.2 | $30.7 | |
| 04 | Science | % Wrap from similar studies | $12.5 | $18.5 | $29.7 | $31.8 |
| 05 | Payload Sys. | Subtotal of below | $105.2 | $147.6 | $247.7 | |
| 05.01 | Payload Sys. Mgmt. | % Wrap from similar studies | $1.68 | $2.36 | $3.96 | |
| 05.02 | Payload Sys. Eng. | % Wrap from similar studies | $1.37 | $1.92 | $3.22 | |
| 05.04 | Optical Instrument | Instrument ROT | $75.49 | $116.67 | $213.89 | |
| 05.05 | Telescope | Stahl Model | $26.64 | $26.64 | $26.64 | |
| 06 | Spacecraft Sys. | $/kg from similar studies | $140.1 | $216.4 | $335.8 | |
| 07 | MOS | % Wrap from similar studies | $18.6 | $18.6 | $18.6 | $16.4 |
| 08 | LVS | AO provided | | | | $150.0 |
| 09 | GDS | % Wrap from similar studies | $19.5 | $19.5 | $19.5 | $6.7 |
| 10 | Project Sys. I&T | % Wrap from similar studies | $15.1 | $22.4 | $35.8 | |
| | Reserves | % Wrap from similar studies | $103.71 | $148.40 | $231.01 | $8.88 |
| | Total | Total of above | $449.4 | $643.1 | $1,001.0 | $218.1 |

| | | | | | |
|---|---|---|---|---|---|
| | | Total A-F | $667.5 | $861.2 | $1,219.1 |
| | | Cost Target (incl LV) | $1,000.0 | $1,000.0 | $1,000.0 |
| A-D Reserves | 30% | Difference | $332.5 | $138.8 | -$219.1 |
| E-F Reserves | 15% | | | | |
| | | **70th percentile cost** | | **$975.7** | |

Table 7. *ATLAS* Probe ROM cost estimate in $M.

Meyer, R.D., et al., 2004, SPIE. 5492, 200, "RITMOS: a micromirror-based multi-object spectrometer"

More, S., Miyatake, H., Takada, M., et al. 2016, ApJ, 825, 39, "Detection of the Splashback Radius and Halo Assembly Bias of Massive Galaxy Clusters"

Moster B. P., Naab T., White S. D. M., 2013, MNRAS 428, 3121, "Galactic star formation and accretion histories from matching galaxies to dark matter haloes"

Parker, A. et al., 2016, PASP, 128, 8010, "Physical Characterization of TNOs with the James Webb Space Telescope"

Perlmutter, S., et al. 1999, ApJ, 517, 565, "Measurements of  and  from 42 High-Redshift Supernovae"

Pisani, A., et al., 2019, "Cosmic voids, a novel probe to shed light on our Universe", Astro2020 science white paper, arXiv:1903.05161

Quijada, M., et al. 2016, SPIE, 9912, 99125V-1, "Optical evaluation of digital micromirror devices (DMDs) with UV-grade fused silica, sapphire, and magnesium fluoride windows and long-term reflectance of bare devices"

Riess, A., et al. 1998, AJ, 116, 1009, "Observational Evidence from Supernovae for an Accelerating Universe and a Cosmological Constant"

Robberto, M.; et al. 2016, Proceedings of the SPIE, 9908, 99088, "SAMOS: a versatile multi-object-spectrograph for the GLAO system SAM at SOAR"

Samushia, L.,  et al.,  2019, in preparation

Seo, H., Eisenstein, D., 2003, ApJ, 598, 720, "Probing Dark Energy with Baryonic Acoustic Oscillations from Future Large Galaxy Redshift Surveys"

Spergel, D.N., et al., 2015, WFIRST SDT Final Report, arXiv:1503.03757

Sullivan, M. et al. 2010, MNRAS, 406, 782, "The dependence of Type Ia Supernovae luminosities on their host galaxies"

Tilvi, V., et al. 2014, ApJ, 794, 5, "Rapid Decline of Ly$\alpha$ Emission toward the Reionization Era"

Tojeiro, R., et al., 2017,  MNRAS, 470, 3720, "Galaxy and Mass Assembly (GAMA): halo formation times and halo assembly bias on the cosmic web"

Travinsky, A. et al. 2016, Optical Eng. 55(9), 094107, "Effects of heavy ion radiation on digital micromirror device performance"

Travinsky, A.; et al., 2017, J. of Astronomical Telescopes, Instruments, and Systems, 3(3), 035003, "Evaluation of digital micromirror devices for use in space-based multiobject spectrometer application"
12